\documentclass{IEEEcsmag}
\pdfoutput=1
\usepackage[colorlinks,allcolors=blue]{hyperref}
\usepackage{enumerate}
\usepackage{lipsum}
\usepackage{siunitx}
\usepackage{todonotes}
\usepackage{upmath}
\usepackage{xifthen}
\usepackage{xspace}

\newcommand{\EMAIL}[1]{\href{mailto:#1}{#1}}

\newcommand{\PKG}[1]{\texttt{\small #1}}
\newcommand{\CRATE}[1]{\href{https://crates.io/crates/#1}{\PKG{#1}}}
\newcommand{\CMD}[1]{\texttt{\small #1}}
\newcommand{\CLI}[1]{\texttt{\small #1}}

\jvol{X}
\jnum{Y}
\paper{Z}
\jmonth{October}
\jname{IEEE Software}
\pubyear{2026}
 
\def\dataCmdCount{102\xspace}
\def\dataGnuBinInstSize{6.7\,MiB\xspace}        \def\dataMultiCallBinSize{14\,MiB\xspace}  \def\dataSingleCallBinSize{172\,MiB\xspace}  

\def\GnuVersion{9.9\xspace}
\def\UutilsVersion{0.4.0\xspace}
\def\ReplicationPackage{\url{https://doi.org/10.5281/zenodo.18735186}}

\begin{document}

\title{Rust Coreutils: Rebuilding Unix Foundations in a Modern Language\textsuperscript{*}}
\author{Sylvestre Ledru}
\affil{Debian, Paris, France}

\author{Samuel Tardieu}
\affil{{LTCI}, Télécom Paris, Institut Polytechnique de Paris, Palaiseau, France}

\author{Stefano Zacchiroli}
\affil{{LTCI}, Télécom Paris, Institut Polytechnique de Paris, Palaiseau, France}

\begin{abstract}
  GNU core utilities (\emph{coreutils}) is a crucial package in modern UNIX systems.
  It comprises around 100 fundamental commands---like \texttt{ls}, \texttt{cp}, and \texttt{cat}---which run every day on millions of computers.
  However, GNU coreutils is also legacy software, with its C codebase dating back to the early 1990s and arguably feature-complete.
  If one were to consider \emph{reimplementing} this essential package, \emph{how} would they do so effectively, and \emph{why}?

  This paper recounts the development of \emph{Rust coreutils}, a contemporary open source reimplementation of GNU coreutils in the Rust programming language, which has reached the status of a drop-in replacement for GNU coreutils, compatible with most Linux distributions.
  By comparing Rust coreutils with its ancestor, we offer insights into creating a reliable substitute for critical software and highlight how modern programming features can attract development interest in legacy packages.
\end{abstract}

 \maketitle

\renewcommand{\thefootnote}{\fnsymbol{footnote}}
\footnotetext[1]{This is the authors' preprint version of IEEE Software article with DOI \href{http://dx.doi.org/10.1109/MS.2026.3733266}{10.1109/MS.2026.3733266}}
\renewcommand{\thefootnote}{\arabic{footnote}}

In the 1970s, Ken Thompson and Dennis Ritchie at Bell Labs developed the Unix operating system with a revolutionary design philosophy: small, modular utility programs---called \emph{coreutils} in this paper---that could be combined through pipes to perform complex tasks~\cite{kernighan2021interview}.
Coreutils include fundamental tools like \CMD{cat}, \CMD{grep}, and \CMD{sort} and form the backbone of the Unix command-line interface, providing users with basic capabilities for file, shell, and text manipulation. Initially implemented in assembly, the original Unix coreutils were ported to C starting in 1972. After that, their codebase remained remarkably similar to its early iterations for decades~\cite{spinellis2021unixevol}.

Over time, alternative coreutils implementations emerged for other Unix-like systems,
each offering its unique take on these foundational tools.
Started in 1990 by David MacKenzie, quickly joined by Jim Meyering, \emph{GNU coreutils}~\cite{mackenzie1994gnucoreutils} aims to provide a compatible replacement for traditional Unix utilities, copyleft licensed in the spirit of the GNU project.
The widespread adoption of GNU+Linux operating systems has led to GNU coreutils becoming the most popular and widely utilized coreutils implementation, serving as a \emph{de facto} gold standard.

In August 2013, Jordi Boggiano started the \emph{uutils} project\footnote{\url{https://uutils.github.io/}} with the goal of reimplementing ``ubiquitous command line utilities'' in Rust, a young programming language at the time: Mozilla released Rust 0.1 in January 2012, reaching 1.0 only in May 2015.
The initial goal was not only to leverage Rust's promise of safety and performance, but also to create a modern cross-platform coreutils implementation that could drop-in replace GNU coreutils.

Note that memory unsafety in GNU coreutils---a common motivation today for ``Rewrite-It-In-Rust'' (RIIR) efforts---was \emph{not} a factor in the creation of uutils. Indeed, GNU coreutils has an excellent security record: 11 CVEs in total over its long history, only two of which were caused by memory-safety bugs.

Uutils was released under the permissive MIT license, the most widely used license within the Rust ecosystem.
This choice would later play a significant role in the adoption of uutils.

The project saw a steady level of activity from 2013 to 2020, with contributors gradually growing the set of available commands and their compatibility with GNU.
With the rising popularity and maturity of Rust, the project gained significant momentum.
Uutils coreutils are now production-ready, and major Linux distributions like Ubuntu are beginning to replace GNU coreutils with them.

In this article we critically recount the development of uutils, comparing them with GNU coreutils from angles including: design, testing, community, and non-functional requirements.
From the uutils experience we distill lessons learned on how to replace foundational software components of popular operating systems and how modern language features can reignite community interest in legacy software.

\section{THE DESIGN OF RUST COREUTILS}

\subsection{Principles}

The main goal of the uutils project evolved over time, converging to \emph{become the default coreutils implementation in major Linux distributions}, a role in the Unix software stack currently fulfilled by GNU coreutils.
Their use in distributions is so widespread that it is hard to track.
In Debian, for example, the (GNU) \PKG{coreutils} package is marked as ``essential'', meaning that other packages can use its commands without declaring an explicit dependency: finding all usages would require runtime testing, with no completeness guarantees.
If you cannot track client code, you cannot adapt it beforehand to behavior changes.
Breakages would not be acceptable in such foundational software: users expect backward compatibility at all costs.
As a consequence, the first design principle is:
\begin{enumerate}[\bfseries P1.]
\item uutils coreutils must be a \textbf{functional drop-in replacement} for GNU coreutils.
\end{enumerate}
which is interpreted by the project as the following functional requirement:
\begin{quote}
  Every successful GNU coreutils command invocation must: 1) succeed with uutils coreutils, 2) return the same exit code, 3) produce identical outputs on \texttt{stdout} and file system.\end{quote}

\smallskip\noindent
While apparently very strict, this requirement allows one to innovate in two places.
First, since invocations that fail with GNU coreutils may succeed with uutils, \emph{new command-line options} can be added to implement new gated features, without breaking backward compatibility.
For example, uutils adds a \CLI{--progress} option to \CMD{cp} and \CMD{mv} that can take significant time when processing many files, providing users with real-time feedback on operation progress.

Second, as divergences on \texttt{stderr} are allowed, it is possible to improve user experience via better, \emph{more informative error messages}.
For instance, when encountering an unrecognized option, GNU \CMD{ls} reports ``\CLI{unrecognized option '--colour'}'', while uutils helps users with the suggestion ``\CLI{error: unexpected argument '--colour' found; tip: a similar argument exists: '--color'}''.

\smallskip
Uutils also aims to leverage features of the Rust language and ecosystem to go beyond what is possible (or easy) in C.
The following design principles derive from this:\begin{enumerate}[\bfseries P1.]\addtocounter{enumi}{1}
\item \textbf{Implement features that are commonplace in modern software}, easy to implement in Rust, and not implemented in GNU coreutils (likely because they are too cumbersome to implement in C).
\end{enumerate}
Examples of uutils features originating from this principle are: native parallelism, Unicode support, and zero-copy data transfer using Linux's \texttt{splice} system call for efficient file operations in commands like \CMD{cat} and \CMD{wc}.

\begin{enumerate}[\bfseries P1.]\addtocounter{enumi}{2}
\item \textbf{Keep the codebase small and maintainable} by resisting NIH (not invented here) syndrome and leveraging the rich ecosystem of existing open source Rust libraries (``crates'') instead.
\end{enumerate}
Depending on more external code comes with drawbacks too, which we analyze and discuss below.

\subsection{Architecture}

The software architecture of uutils is star-shaped.
The source tree contains multiple software packages, one for each of the \dataCmdCount implemented coreutils binaries (e.g., \CMD{stat}, \CMD{cut}, etc.), all depending on a shared \CRATE{uucore} crate that implements helper functions to manipulate the filesystem, locales, processes, errors, etc. This is similar to GNU coreutils, where Gnulib contains shared code.
A notable difference is that the latter contains a large amount of code to implement common data structures and portability functionalities; \CRATE{uucore} does not need to, as equivalent functionalities are provided by the Rust standard library.
The Rust-native \PKG{Cargo} build system is used to decide what to build, in terms of both binary selection and conditional compilation for OS portability. 

Differently from C, the Rust compiler defaults to static linking (of Rust code).
Hence, installing one separate binary per coreutils command would result in significant duplication: the code of \CRATE{uucore} and all Rust coreutils transitive dependencies---which abound, due to design principle P3---would be copied 100+ times on the target system, for an installed size of \dataSingleCallBinSize.
To avoid this, uutils, similarly to BusyBox, installs a single multi-call \CMD{coreutils} executable~\cite{ibm2002multicall}, with a collection of links pointing to it, and a dispatcher that decides what to do based on \texttt{argv[0]}.
This reduces the code size to \dataMultiCallBinSize; this is up from \dataGnuBinInstSize for GNU, but still acceptable even for resource-constrained systems.

\section{TESTING UNIX CLI FOUNDATIONS}

\subsection{End-to-end testing}

Testing uutils poses two challenges.
First, there is no rigorous specification of \emph{GNU} coreutils, which uutils aims to replace.
The POSIX standard~\cite[Utilities]{ieee2024posix} specifies only a small subset of the coreutils commands, and in any case uutils must replicate the \emph{divergences}\footnote{``The GNU utilities documented here are \emph{mostly} compatible with the POSIX standard.''~\cite[Introduction]{mackenzie1994gnucoreutils} (emphasis added)} of GNU coreutils from the standard, to respect principle P1.

To address this difficulty, uutils employs a three-tier testing strategy:
\begin{enumerate}
\item \textbf{Unit tests}: Function-level validation using Rust native testing framework for fast developer feedback.
\item \textbf{Integration tests}: Over \num{4200} tests executing in under 40 seconds, covering both common usage patterns and edge cases, enabling rapid local testing.
\item \textbf{External end-to-end test suites}: Running GNU coreutils (as well as Toybox and BusyBox) test suites to verify cross-implementation compatibility.
  External test suites serve as the ultimate benchmark for drop-in compatibility.
\end{enumerate}
All uutils commands run through the GNU coreutils end-to-end tests for each CI run: they receive the same inputs and must produce the same outputs (except on \texttt{stderr}, as discussed) for a test case to succeed.
After each GNU coreutils release, uutils updates its test references and adjusts either the tests or implementation code to maintain compatibility.
Testing against Toybox and BusyBox provides additional validation, exercising different code paths and revealing assumptions that might not surface when testing against a single external reference.

\begin{figure*}
  \includegraphics[width=\linewidth]{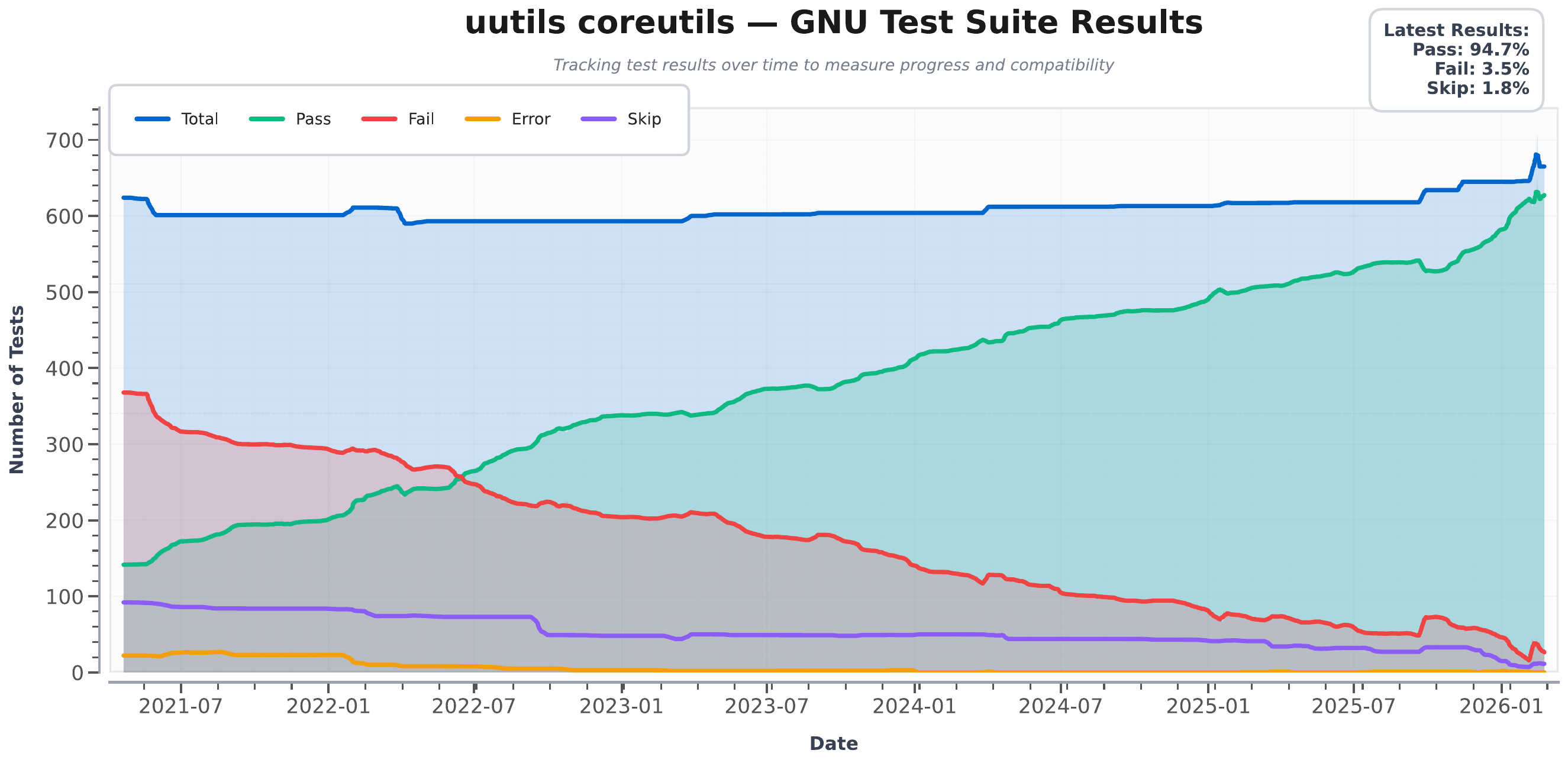}
  \caption{Results over time of uutils end-to-end testing on the GNU coreutils test suite.}
  \label{fig:difftesting-gnu-rust}
\end{figure*}

The GNU test suite is extensive: 630+ test cases, implemented in 25\,kSLOC of scripting languages.
Figure~\ref{fig:difftesting-gnu-rust} shows the evolution over time of how uutils fares on it.
Over time the success rate of uutils increased linearly, almost closing the gap with GNU.
Remaining test failures today are due to lack of prioritization on some unpopular programs or weird edge cases that do not impact day-to-day usage.

This approach also created gamification opportunities for uutils development.
By monitoring the evolution over time and breaking it down by utility,\footnote{\url{https://uutils.github.io/coreutils/docs/test_coverage.html\#coverage-per-category}} contributors are motivated to reduce the gap with GNU and feel rewarded when their patches make a dent.

\subsection{Differential testing and fuzzing}

The second testing difficulty for uutils is that coreutils programs tend to have a lot of options.
Depending on their order, combinations, and validity, program behavior might change.
Despite the good properties of all discussed test suites, this CLI combinatorics makes it unfeasible to test all cases, even before worrying about the testing space of file-based inputs.

To start addressing this problem, uutils uses traditional fuzzing, which is historically very fitting, as the seminal fuzzing experiment by Miller et al.~\cite{miller1990unixfuzzing} in 1990 consisted in throwing randomly generated inputs to CLI UNIX utilities, including many coreutils commands.
At the time, they were able to crash 24\% of the tested commands.
Today, uutils benefits from integration with Google's OSS-Fuzz infrastructure~\cite{serebryany2017ossfuzz}, which provides fuzzing resources to selected open source projects for free.
The fuzzing setup is also easily accessible to contributors: anyone can run fuzzing locally with simple commands, making it a first-class testing tool in the development workflow.

To further strengthen testing, uutils also deploys differential testing~\cite{mckeeman1998difftesting} which, we recall, is an adaptation of fuzzing to contexts where multiple implementations of the same application exist and should behave the same.
In differential testing, the test runner generates a random input, feeds it to multiple implementations, and uses them as cross-oracles: if outputs differ, a failure is reported.
This context fits the case of coreutils perfectly.

Inspired by recent results on browser JavaScript engines~\cite{lima2021javascriptdifftesting}, uutils uses differential testing to compare its CLI behavior with that of GNU.
Input generation is guided by grammars that encode the synopsis of coreutils commands (so that the time spent trying invalid inputs is minimal).
LibFuzzer~\cite{serebryany2016libfuzzer} guides the actual fuzzing phase, based on code coverage to discard fuzzing seeds that do not lead to unexplored code paths, and also reduces failure-inducing inputs, to minimize the size of newly discovered test cases.
As with the traditional test suites, output comparison focuses on exit code and
stdout, ignoring stderr divergences.

The project's experience with differential fuzzing proved valuable as it revealed gaps not only in the uutils implementation, but also in GNU coreutils' test coverage, leading to improved testing for \emph{both} projects.
It identified missing test cases in the GNU test suite, prompting additions that strengthen the overall robustness of GNU coreutils.
For example, differential fuzzing uncovered discrepancies in how the two implementations handle timezone specifications in date strings.
Running \CLI{TZ=UTC0 date -u -d TZ="UTC+5:30" 2025-01-01} produces \CLI{Wed Jan  1 04:30:00 +05:30 2025} in uutils but \CLI{Wed Jan  1 05:30:00 UTC 2025} in GNU, revealing edge cases that were not covered by test suites.
Another example is the \CMD{realpath} command, where differential fuzzing identified subtle differences in path resolution implementations, leading to improved test coverage for both projects.

The collaboration between uutils and GNU coreutils has been excellent throughout this process.
Uutils contributors have provided many test cases to improve the GNU test suite, while GNU coreutils maintainers have actively participated by reporting bugs against uutils, adjusting GNU implementations (such as the \CMD{cksum} command), and contributing to ongoing discussions.

\section{COREUTILS COMPARISON}

GNU coreutils and uutils differ in relevant ways.
To get a grasp of their differences, we compare the two projects under the following aspects of the software development process: dependencies, complexity, and repository activity.
All reported data are as of GNU coreutils version \GnuVersion and uutils \UutilsVersion. Measurements can be reproduced using the reproducibility package available at \ReplicationPackage.

\subsection{Dependencies and supply chain}

\begin{table}
  \caption{Number of transitive dependencies in uutils in the final executables (runtime) and used during build and testing (runtime+dev).}
  \centering
  \begin{tabular}{l|r|r}
    \textbf{Dependency~depth} & \textbf{Runtime} & \textbf{Runtime + dev} \\
    \hline
    1 (direct deps.) & 8 & 31 \\
    2 & 29 & 97 \\
    3 & 55 & 153 \\
    4 & 84 & 201 \\
    5 & 109 & 210 \\
    6 (max depth) & 110 & 213
  \end{tabular}
  \label{tab:cargo-tree}
\end{table}

The GNU coreutils codebase is predominantly self-contained, with its primary external dependencies being the GNU portability library (Gnulib) and essential build tools such as Bison for parsing and Gettext for internationalization.
Uutils in contrast, according to principle P3 and to implement P2 more easily, leverages the Rust ecosystem to avoid redundant implementations of higher-level functionalities like output coloring and CLI parsing, as well as streamline development.

Table~\ref{tab:cargo-tree} shows the number of (transitive) dependencies of uutils.
With optional features enabled, the 8 direct dependencies of uutils grow to 110 external crates included in the final executables distributed to users.
Including development dependencies, up to 213 external crates are part of the uutils supply chain.

GNU coreutils only list 15 tools used for development (in addition to Gnulib) with only Git, Perl, and XZ Utils originating from outside the GNU project.

Incorporating Gnulib expands the GNU coreutils codebase from approximately \num{60 000} (non-blank, non-comment) lines of code to nearly \num{200 000} overall: a technical leverage~\cite{massacci2021techleverage} ratio of $\times3.3$.
For uutils, the inclusion of dependencies increases the codebase from around \num{90 000} to over 3 million lines: $\times33.3$ leverage, 10 times higher than GNU coreutils.
While not all of this external code is actively utilized in either project, depending on it increases exposure to supply chain attacks~\cite{ladisa2023soksupplychain}.

GNU coreutils, by primarily depending on GNU code arguably requires less scrutiny than uutils.
In the latter, even a minor, blind update to a dependency could introduce unvetted or potentially malicious code, necessitating heightened vigilance in dependency management.
To mitigate this risk, uutils vets external dependencies based on common criteria (popularity, license, etc.),
and actively avoids introducing dependencies to young crates maintained by a single person.
If needed, the project also forks crates abandoned upstream to take over their maintenance, as it happened for \CRATE{uutils\_term\_grid} and \CRATE{ansi-width}.

\subsection{Code complexity}

A comparative analysis of the two codebases using the Lizard code complexity analyzer\footnote{\url{https://github.com/terryyin/lizard}, accessed 2026-02-23}
reveals a notable disparity in cyclomatic complexity~\cite{1702388}: GNU coreutils functions exhibit an average complexity of 9.40; uutils 3.30.
This discrepancy is due to several factors.\footnote{The \texttt{rust-code-analysis-cli} complexity analyzer~\cite{ARDITO2020100635} yields broadly similar results. In addition, its cognitive complexity results are consistent with the cyclomatic complexity measurements.}

Both projects prioritize robust error handling.
In C, thorough error checking is cumbersome, as it necessitates explicit validation of return values before propagating errors to the calling function, often requiring manual resource cleanup, such as deallocating temporary memory.
Rust simplifies error handling with the \texttt{?} postfix operator, which automatically propagates errors or the absence of values to the caller, while destructors ensure automatic resource cleanup. Lizard reports how complex the code looks to the programmer: destructors are hidden and do not count toward the reported complexity, even though complex control flow may be invoked behind the scenes.

The uutils codebase emphasizes simplicity and readability (principle P3) by leveraging higher-level language constructs, such as iterators and pattern matching, native to Rust.
On the other hand, GNU coreutils relies on lower-level constructs like loops and lengthy \texttt{if-else} chains, particularly when comparing strings against multiple possible values.
Uutils also favors a modular approach, in which single-purpose functions are factored out in common libraries, similar to observed evolution trends in Unix system architectures~\cite{spinellis2021unixevol}.
This design choice results in a larger number of functions in uutils---over 2800---compared to GNU coreutils, which comprises approximately 1400 functions.

Additionally, CLI option handling in uutils is implemented with the \CRATE{clap} crate, which automates option parsing, validates arguments, checks for incompatibilities, generates help and error messages, and manages type conversions.
In contrast, while GNU coreutils reuse \texttt{getopt\_long()} for low-level option parsing, higher level functionalities like validation, type conversions, and message generation are independently reimplemented in each tool.

Taken together, uutils' choices enhance maintainability by keeping boilerplate to a minimum and reducing the overall code complexity.

\subsection{Repository activity}

\begin{figure*}
  \includegraphics[width=\linewidth]{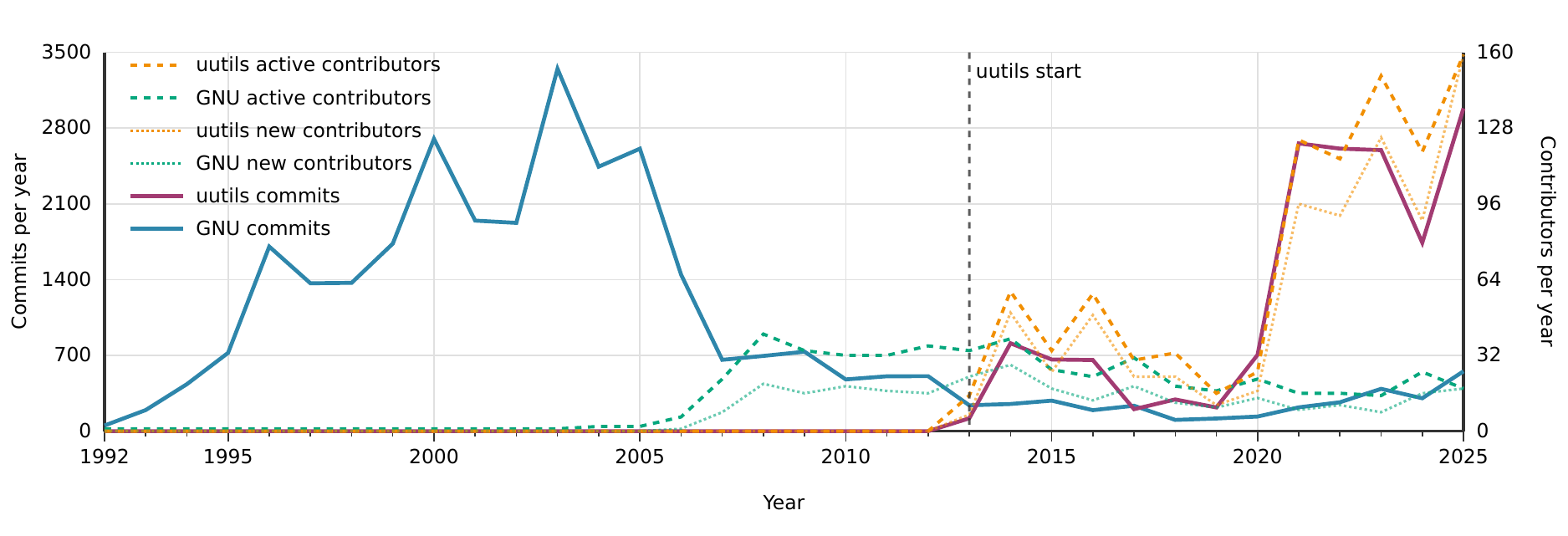}
  \caption{Comparison of development activity in GNU coreutils vs.~uutils, under various metrics.
Note that year 2025 was still incomplete at measurement time, by about 1.5 months.}
  \label{fig:activity-comparison}
\end{figure*}

Figure~\ref{fig:activity-comparison} shows the yearly activity in the main Git repositories of GNU coreutils and uutils, excluding bots.
From the early 90s to 2005, GNU coreutils was a very active project, with thousands of commits per year, made primarily by Jim Meyering, the project maintainer.
After 2005 the project activity decreased, reaching a steady state, typical of maintained, stable projects.
The number of GNU coreutils yearly contributors increased since the early days of the project, then decreased over the last decade, both in terms of active and new contributors (with no commits before that year).

Uutils history before 2020 shows moderate activity, with 1000 commits/year authored by about 50 yearly contributors, many of whom new.
Starting from 2020 there is a clear spike of activity, still growing today, with 2500 commits/year and 125+ active contributors, with a continued stream of new contributors every year.

Contribution concentration in GNU coreutils is very high: the top 10 contributors account for nearly 98\% of all commits, whereas in uutils the top 10 contributors are responsible for less than 65\% of the total commits---a healthier ``bus factor''.

The uutils project also reflects broader trends in systems programming demographics and maintainability.
As the pool of experienced C developers gradually decreases, maintaining critical infrastructure written in C will become increasingly challenging in the long run.
Rust modern tooling, built-in memory safety, and active ecosystem lower the barrier to entry for younger contributors.
The sustained contributor growth shown in Figure~\ref{fig:activity-comparison} suggests that reimplementing foundational utilities in modern languages can attract developer interest that might otherwise be directed elsewhere, potentially addressing long-term maintenance concerns for essential system software.

\section{OPERATING SYSTEM INTEGRATION}

The adoption of uutils progressed from experimental toy usage by hobbyists to massive adoption by major Linux-based operating systems (OS).
Apertis, a Linux distribution for the automotive sector, was one of the earliest OS adopters to replace GNU coreutils with uutils in their \texttt{v2022dev3} release, to avoid GPL version 3 licensing restrictions while retaining backward compatibility.
Snap Spectacles, an OS for smart glasses, did the same in September 2024.

A significant adoption milestone was achieved with Ubuntu 25.10 (late 2025), becoming the first general-purpose Linux distribution to ship uutils as default coreutils to an estimated user base in the tens of millions.
Testing in development conditions, like uutils already does, provides no guarantee of covering all workflows of a user base like Ubuntu's.
Nothing short of deploying to all users and collecting their feedback can reliably identify all edge cases and compatibility issues.
Indeed, Ubuntu adoption yielded invaluable feedback and uncovered issues not caught earlier on.

One significant bug affected the \PKG{unattended-upgrade} system, where subtle differences in the date command caused upgrade failures in unrelated packages.\footnote{\url{https://bugs.launchpad.net/ubuntu/+source/rust-coreutils/+bug/2127970}} Another interesting case involved the use of \CMD{dd} by the \PKG{makeself} package, revealing edge cases in input validation and error handling that only manifested in specific scripting contexts.\footnote{\url{https://bugs.launchpad.net/ubuntu/+source/makeself/+bug/2125535}}
A comprehensive list of issues reported through Ubuntu can be found in the uutils GitHub repository under the ``reported-canonical'' and ``reported-launchpad'' labels.\footnote{\url{https://github.com/uutils/coreutils/issues?q=label\%3A\%22reported-canonical\%22} and \url{https://github.com/uutils/coreutils/issues?q=label\%3Areported-launchpad}}

The Ubuntu deployment also highlighted performance regressions not caught by micro benchmarks, but significant in real-world usage scenarios.
This led to the introduction in uutils of a comprehensive benchmarking framework using \href{https://codspeed.io}{CodSpeed}, that provides continuous performance monitoring across releases to prevent future regressions from reaching production users.

As a last resort mechanism to mitigate the impact of compatibility issues during the transition, and react quickly in case of need, Ubuntu also implemented a mechanism that allows users to switch back to GNU coreutils, on a per-machine basis, while retaining uutils as default for all installations.

Overall, Ubuntu adoption demonstrates a successful transition, with some caveats.
While last-minute compatibility issues emerged and required prompt fixes, the integration achieved its main goal of validating uutils production-use at scale.
The identified bugs, though annoying for affected workflows, represented only a small fraction of users and were resolved quickly without widespread user disruption nor aborting the migration.
Most importantly, the deployment proved that coreutils migration is achievable, and now provides a roadmap for other distributions.

\section{LESSONS LEARNED AND OUTLOOK}

The story behind uutils is instructive for several stakeholders in the software engineering community.
It complements well that of previous C-to-C kernel reimplementations (e.g., Cygwin, Wine, FreeDOS) from the vantage point of core Unix executables and in a cross-language context.

The first takeaway is that it is possible to migrate a software component so entrenched and with countless unseen usage patterns as coreutils, from one implementation to another.
It is even possible to do so at the scale of tens of millions of deployments, like Ubuntu, in the relatively short time frame of a few years.
Doing so requires some discipline, in particular for testing: strict design principles about compatibility, differential testing and fuzzing, systematic monitoring, and large-scale user testing were all key ingredients of uutils' success at this task.

The second takeaway is that new technology (Rust, in this case) can reignite interest in ``old'' technology (coreutils) that had previously been considered legacy/feature-complete for many years prior.
Aside from the interest from tech media, the tangible amount of development activity in the repository shows that uutils is both a healthy and attractive project for generations of developers that have not considered contributing to GNU coreutils before.
Along the way, the uutils community also discovered that GNU coreutils was not as feature complete as everyone thought, resulting in new, useful features that are now being implemented in \emph{both} projects, benefiting everyone.

Finally, uutils is a paradigmatic example of a well-functioning open source ``funnel'' of contributor pathways: many initial drive-by contributors, some of whom remain to become regular contributors, some of whom end up becoming project maintainers.
To support this model, uutils leveraged the set of initially failing GNU coreutils compatibility tests as a pool of easy tasks for newcomers.
This led to some sort of gamification, pushing the community forward together and attracting contributors.

\medskip

Zooming out a bit, uutils is part of a larger trend in the tech sector: system-level code bases are migrating from C/C++ to Rust.
At lower abstraction levels: Linux now supports kernel code written in Rust, the first system-level programming language other than C to be used for Linux kernel development in over 30 years.
At higher abstraction levels: Rust is everywhere from compilers to web browsers, from package managers to web frameworks, to the point that the DARPA agency of the US department of defense publicly advertises an ambitious program to semi-automatically ``[Translate] All C to Rust''.\footnote{\url{https://www.darpa.mil/research/programs/translating-all-c-to-rust}, accessed 2025-12-08}

Different stakeholders are getting into Rust for different reasons (security, performances, zero-cost abstractions, ecosystem) and analyzing the growing popularity of Rust is beyond the scope of this article.
Still, in this context uutils remains a telling case study of how one can swap a fundamental UNIX component, retaining backward compatibility and benefiting from the features that a modern language has to offer to system programmers.

\medskip

Moving forward, the uutils project will be busy.
On one hand, with this level of adoption comes greater exposure and maintenance demands.
As uutils embraced innovation in user-facing features, it is likely to face an increased demand for these features---something that GNU coreutils experienced to a lesser degree due to its long-standing status as feature-complete.

On the other hand, uutils has increased its scope and is now targeting the Rust reimplementation of diffutils, findutils, as well as the \texttt{sed} command line editor~\cite{spinellis2025sedrust}.
For system-level programmers looking for new challenges, this looks like the perfect time to get involved and leave their mark in open source code bases that might become the new foundations of our digital infrastructure.
For empirical researchers, they will soon have a treasure trove of data about a new and growing ecosystem of system-level applications to extract insights from.

\section{ACKNOWLEDGMENTS}

The authors would like to thank Jordi Boggiano, founder of the uutils project, as well as the $\approx\,$700 contributors to it\ldots thus far!

\begin{IEEEbiography}{Sylvestre Ledru}{\,}is the lead of the uutils project.
  He focuses on open-source software development, web browsers, operating system design, and compilers.
  He has been contributing to Debian for 20 years, focusing on large-scale changes such as rebuilding the archive with Clang and maintaining compiler toolchains including LLVM/Clang and Rust.

  Contact: \EMAIL{sylvestre@debian.org}.
\end{IEEEbiography}

\begin{IEEEbiography}{Samuel Tardieu}{\,}leads the Autonomous Critical Embedded Systems team at Télécom Paris, Polytechnic Institute of Paris. He focuses on developing safe, secure, and frugal critical systems, from their design through to their runtime behavior. He actively contributes to the development and maintenance of the Clippy linter for Rust.

  Contact: \EMAIL{samuel.tardieu@telecom-paris.fr}.
\end{IEEEbiography}

\begin{IEEEbiography}{Stefano Zacchiroli}{\,}is full professor of computer science at Télécom Paris, Polytechnic Institute of Paris.
  His current research interests span digital commons, open source software engineering, computer security, and the software supply chain.
  He is co-founder and CSO of Software Heritage, the largest public archive of software source code.

  Contact: \EMAIL{stefano.zacchiroli@telecom-paris.fr}.
\end{IEEEbiography}

\end{document}